**Transcutaneous auricular vagus nerve stimulation modulates auditory-evoked responses and conflict adaptation in humans**

**Issey Kamatani[1#], Yasuo Kabe[1#], Shinichi Kumagai[1,2#], Yoshiyuki Onuki[2], Hexin Xu[1], Tomoyo Isoguchi Shiramatsu[1], Naoto Kunii[2], Kensuke Kawai[2] and Hirokazu Takahashi[1*]**

[1]Graduate School of Information Science and Technology, The University of Tokyo, Tokyo, Japan; [2]Department of Neurosurgery, Jichi Medical University, Shimotsuke, Japan

# These authors contributed equally to the present work.

***Correspondence:**

Hirokazu Takahashi, Ph.D., Department of Mechano-Informatics, Graduate School of Information Science and Technology, The University of Tokyo, 7-3-1 Hongo, Bunkyo, Tokyo 113-8656, Japan. Tel: +81-3-5841-0461. Email: takahashi@i.u-tokyo.ac.jp

**Abstract**

Vagus nerve stimulation (VNS) engages ascending neuromodulatory systems, and animal studies have motivated the hypothesis that it shifts sensory cortical processing toward feedforward (FF) relative to feedback (FB) transmission. Whether transcutaneous auricular VNS (taVNS) produces a corresponding pattern in humans remains unclear. In a single-blind, sham-controlled crossover study, 11 healthy adults received 120 min of taVNS (25 Hz, left tragus) and sham stimulation (left earlobe) in separate sessions. Click-evoked EEG was recorded before, twice during, and immediately after stimulation. Sensory discrimination thresholds and conflict adaptation in a Simon task were assessed before and during stimulation. Relative to Sham, taVNS attenuated the decline in the early P0 response across the session and enhanced the Pa–Nb response early during stimulation. The later P2 response did not show the increase observed under Sham. After stimulation, taVNS reduced click-locked theta (5–8 Hz) inter-trial phase coherence relative to Sham, with no significant between-condition difference in the change in theta power. Conflict adaptation increased under taVNS, whereas sensory thresholds and the overall Simon effect did not differ significantly between conditions. In exploratory analysis, greater decreases in theta phase coherence were associated with greater increases in conflict adaptation. These findings indicate that taVNS modulates auditory responses across processing stages and conflict-related behavior. The pattern is compatible with altered FF/FB weighting, but the direction of information flow and the underlying neuromodulatory mechanisms were not measured directly.

## 1 Introduction

Vagus nerve stimulation (VNS) is used clinically to treat drug-resistant epilepsy and treatment-resistant depression (Englot et al., 2011; Kawai et al., 2017; Rush et al., 2005). VNS engages cholinergic and noradrenergic systems that alter cortical responses (Hulsey et al., 2017; Collins et al., 2021; Mridha et al., 2021; Bowles et al., 2022; Martin et al., 2024), but how it changes sensory information processing remains poorly understood. In rats, VNS preferentially enhanced sound-evoked responses in superficial auditory cortical layers and increased evoked gamma activity while reducing evoked theta activity (Takahashi et al., 2020; Kumagai et al., 2023). Within a predictive coding framework, these observations motivated the hypothesis that VNS increases the weighting of prediction errors conveyed by feedforward (FF) pathways relative to predictions conveyed by feedback (FB) pathways (Kumagai et al., 2025).

Whether transcutaneous auricular VNS (taVNS) in humans produces a corresponding pattern remains unclear. Human studies have mainly examined later auditory event-related responses arising after 100 ms, such as the P3 component, with variable findings (De Taeye et al., 2014; Rufener et al., 2018; Gadeyne et al., 2022; D'Agostini et al., 2023). Because FB contributions increase in this latency range (Garrido et al., 2007), these responses combine FF and FB contributions and cannot, on their own, establish whether the balance shifts toward FF or FB.

Earlier auditory middle-latency responses (MLRs) provide complementary measures of ascending, predominantly FF processing. In rats, early cortical evoked potentials reflect the encoding of basic acoustic features, including frequency and intensity (Takahashi et al., 2004, 2005a, 2005b). Microstimulation of either the cochlear nucleus or the ventral medial geniculate body elicits tonotopically organized cortical activation (Takahashi et al., 2005b; Shiramatsu et al., 2016). Together with their short latencies, these findings support a prominent contribution of ascending sensory transmission to early evoked responses. Early auditory evoked potentials in rats include both subcortical and cortical contributions (Takahashi et al., 2007), as do human MLRs (Picton et al., 1974; Kadoya et al., 1988). This correspondence provides a physiological basis for comparing early auditory responses across species and motivates the use of human MLRs to test for VNS-induced modulation of early FF processing. To our knowledge, the earliest MLR components have not been examined under taVNS.

We therefore focused on two MLR components: P0 (5–15 ms), which is thought to primarily reflect early subcortical auditory activity (Hashimoto, 1982; Kadoya et al., 1988; Leavitt et al., 2007), and the Pa–Nb complex (25–52 ms), which reflects auditory cortical processing (Picton et al., 1974; Yvert et al., 2001; Picton, 2011). If the FF/FB shift proposed on the basis of rat studies (Kumagai et al., 2025) also occurs in humans, these predominantly afferent responses should be larger under taVNS than under sham stimulation, whereas the slow-vertex P2 and click-locked theta-band activity, which may have greater FB contributions (Garrido et al., 2007; Fontolan et al., 2014), should be reduced. At the behavioral level, such a shift should also increase conflict adaptation, the trial-to-trial adjustment of control driven by conflict (Botvinick et al., 2001), which taVNS has been reported to enhance (Fischer et al., 2018). We tested these predictions in healthy adults in a single-blind, sham-controlled crossover study of 120 min of taVNS.

## 2 Materials and Methods

### 2.1 Participants and Design

Eleven healthy adults (3 women; age 24.6 ± 4.2 years (mean ± s.d.), range 21–34) participated after providing written informed consent. The protocol was approved by the ethics committee of Jichi Medical University (clinical research review 23-078) and authorized as a multi-site study by the Graduate School of Information Science and Technology, The University of Tokyo (230809-1), and the study was conducted in accordance with the Declaration of Helsinki. Each participant completed a taVNS session and a sham session in randomized order (taVNS first in four participants) separated by at least 12 h (measured interval: median 23.8 h, range 22.4 h to 43.8 days). Participants were blinded to the stimulation condition (single-blind design). Because of missing or incomplete recordings, sample sizes differed across measures: in one participant, the Early and Post click recordings of the sham session were interrupted after about 5 s (EEG, $N = 10$); in the Simon task, the log file for one participant's taVNS Pre run was not saved and another participant's taVNS During run was stopped after 76 of 192 trials ($N = 9$); the threshold data were complete ($N = 11$). The sample size for each analysis is reported with the corresponding results. The sample size was comparable to previous studies of noninvasive VNS with EEG measures in healthy adults (Lewine et al., 2019; Mertens et al., 2022).

**2.2 Stimulation**

Stimulation was delivered with a Model 0125 stimulator (Soterix Medical, Woodbridge, NJ, USA) through a custom ear-clip electrode with carbon contacts and carbon leads (TK223-005; Unique Medical, Tokyo, Japan), attached to the left tragus for taVNS and to the left earlobe for sham stimulation (Sham) (Figure 1B). The electrode came in two sizes, with circular carbon contacts 8 mm or 3 mm in diameter (approximately 0.50 or 0.07 cm²) within a 1-mm plastic rim (outer diameter 10 or 5 mm); the larger electrode was used unless the participant's tragus was too small to hold it. The tragus receives fibers of the auricular branch of the vagus nerve, and its stimulation at 25 Hz engages central vagal projections more than earlobe stimulation does (Badran et al., 2018), whereas the earlobe is considered to lack vagal innervation (Peuker and Filler, 2002). Monophasic 250-μs pulses were delivered at 25 Hz in 30-s on/30-s off cycles for 120 min. In each session, intensity was titrated from 0.5 mA in 0.2-mA steps to the strongest non-painful tingling sensation (a rating of 8) on a 0–9 scale (0, nothing; 3, weak tingling; 6, strong tingling; 9, pain). The resulting intensity was 2.16 ± 0.74 mA for taVNS and 2.49 ± 0.69 mA for Sham.

**2.3 Session Protocol**

Each session (Figure 1A) comprised four 10-min blocks of 1-Hz clicks and two runs of the behavioral tasks. The first click block (Pre; −80 to −70 min relative to stimulation onset) was followed by the sensory-threshold tasks (10 min) and the Number Simon task (30 min). Stimulation then started. The second click block (Early; 0–10 min) began at stimulation onset. After 40 min without recording, the third click block (Mid; 50–60 min) was recorded, followed by a second run of the behavioral tasks. Stimulation stopped at 120 min, and the fourth click block (Post; 120–130 min) was recorded immediately afterwards. Thus, Early and Mid were recorded during stimulation, and Post was recorded immediately after stimulation ended. The two runs of the behavioral tasks are referred to as Pre (sensory-threshold tasks at about −60 to −50 min and the Simon task at about −40 to −10 min) and During (about 70–80 min and 90–120 min after onset, respectively). In both runs, 1-Hz tone bursts (5 min) and 40-Hz clicks (5 min) were also recorded before the sensory-threshold tasks, and a visual evoked potential paradigm (10 min) was presented before the Simon task; these recordings are not analyzed here.

**2.4 Auditory Stimuli**

Clicks generated using 1-ms monophasic pulses from a multifunction generator (WF1974, NF Corporation, Japan) were presented to the right ear through insert earphones (ER-3C, Etymotic Research, Elk Grove Village, IL, USA) at 1 Hz (600 clicks per 10-min block). The sound level was not calibrated in dB SPL; it was set to the loudest level that participants reported as comfortable. During recording, participants sat with their eyes open, fixating a cross, and were instructed to ignore the clicks.

**2.5 EEG Acquisition and Preprocessing**

EEG was recorded from 32 scalp electrodes (EasyCap, Easycap GmbH, Wörthsee, Germany; international 10–20 positions, Cz reference, Fpz ground, impedances below 25 kΩ) with an actiCHamp Plus amplifier and BrainVision Recorder (Brain Products GmbH, Gilching, Germany) at 1000 Hz. Preprocessing was performed using EEGLAB (v2024.2; Delorme and Makeig, 2004) in MATLAB (R2025a). Line noise (50 Hz) and stimulator noise (25 Hz and its harmonics) were attenuated with ZapLine-plus (de Cheveigné, 2020; Klug and Kloosterman, 2022), with spectral lines detected between 17 and 99 Hz in each recording. Bad channels were detected with the EEGLAB function pop_rejchan (normalized kurtosis >6 SD; not applied when >15% of channels were flagged) and interpolated with spherical splines (Perrin et al., 1989), as in Arslan et al. (2025). The data were then re-referenced to the common average, with Cz restored. Epochs were rejected when their maximum absolute amplitude across channels and time points exceeded the mean + 5 SD of this measure across all clicks within each recording. This criterion was applied separately to the click-evoked potential and theta analyses. More than 99.4% of epochs were retained at every time point in both conditions.

## 2.6 Stimulator Artifact

Because taVNS was delivered at 25 Hz, stimulation artifacts were assessed in both continuous EEG recordings and click-locked averages. At Cz (common average reference), the 25-Hz spectral peak was quantified relative to the local spectral background using Welch's method in raw and preprocessed recordings. Its scalp distribution was also examined in raw Early and Mid recordings. To assess residual artifacts in the click-locked averages, MLR epochs were averaged with alternating signs (± average; Schimmel, 1967). The 25-Hz amplitude was quantified using a single-frequency discrete Fourier transform over −100 to 500 ms. The 25-Hz and Pa–Nb amplitudes in these averages were compared between conditions using sign-flip permutation tests, without correction for multiple comparisons.

## 2.7 Click-Evoked Potentials

The continuous data were filtered separately for MLRs (10–250 Hz; Shiga et al., 2015; McFadden et al., 2021) and slow-vertex responses (1–30 Hz; Picton, 2011) using zero-phase FIR filters (EEGLAB pop_eegfiltnew, default settings). Epochs from −100 to 500 ms around each click were then extracted and baseline-corrected (−100 to 0 ms). P0, Pa–Nb and P2 were measured from each participant's average waveform. P0 was defined as the maximum amplitude within 5–15 ms, Pa–Nb as the peak-to-trough amplitude between Pa (25–40 ms) and Nb (40–52 ms) (Todd et al., 2014; McFadden et al., 2021), and P2 as the maximum amplitude within 125–175 ms. P0 was measured at Cz, a conventional site for early auditory MLRs (Picton et al., 1974; Picton, 2011; McFadden et al., 2021). P2 and theta measures were assessed over a fronto-central region of interest (ROI; Figure 1C) around FCz, consistent with previous studies of slow-vertex responses (Näätänen and Picton, 1987; Picton, 2011) and auditory theta phase coherence (Ko et al., 2012). Because FCz was not included in the montage, the ROI comprised Fz, FC1, FC2, Cz, C3 and C4, and the signals were averaged across these electrodes before measurement. The ROI was defined on the basis of previous studies, independently of the present data. Pa–Nb, an index of early auditory cortical processing (Picton, 2011), was measured in the left ROI (FC1 and C3), contralateral to the ear receiving the clicks, based on a previous study showing contralateral predominance of early auditory responses to monaural stimulation (Kadoya et al., 1988).

## 2.8 Theta-Band Measures

Theta analyses focused on the Pre-to-Post change because VNS effects on auditory-evoked oscillatory activity became more pronounced over 2 h of stimulation in rats (Kumagai et al., 2023). Inter-trial phase coherence (ITPC) and event-related spectral perturbation (ERSP) were computed with six-cycle Morlet wavelets implemented in NumPy and SciPy. The data were first high-pass filtered at 0.5 Hz (fourth-order Butterworth, zero-phase). To minimize edge artifacts, wavelet transforms were then applied to the continuous recording before epoching. For ERSP, power was averaged across trials and expressed in dB relative to the mean power in a pre-click baseline (−300 to −100 ms) (Makeig, 1993). ITPC and ERSP were averaged over 5–8 Hz. Summary measures were additionally averaged over 0–300 ms after click onset. The lower frequency limit was set at 5 Hz to limit temporal smearing across successive clicks. ITPC and ERSP were computed from signals averaged across ROI electrodes, whereas scalp maps were computed at individual electrodes. As a wavelet-independent check, ITPC was also estimated at Cz using a 1-s discrete Fourier transform of each inter-click interval (Supplementary Figure S3).

**2.9 Behavior**

Visual and auditory discrimination thresholds (Figure 1D) were measured with a QUEST type Bayesian adaptive procedure (Watson and Pelli, 1983), implemented in PsychoPy, with unspeeded yes/no responses. In the visual task (orientation discrimination from vertical; Heeley and Buchanan-Smith, 1990), a sinusoidal grating with a Gaussian envelope (0.0398 cycles/pixel, contrast 0.75) was shown for 300 ms, tilted clockwise from vertical. Participants reported whether they perceived a tilt. In the auditory task (pure-tone frequency discrimination; Creelman and Macmillan, 1979), a 1000-Hz reference tone and a comparison tone (1000 + Δf Hz) were presented sequentially (600 ms each; comparison onset 800 ms after reference onset), and participants reported whether they perceived a pitch difference. In each behavioral assessment, each sensory task was performed twice, each time with two staircases, one starting at a high stimulus level and the other at a low stimulus level (40 trials per staircase for the visual task and 30 for the auditory task). The threshold was estimated by maximum-likelihood fitting of a cumulative Gaussian function of $\log_{10}$ stimulus magnitude to all trials from the four staircases (160 visual and 120 auditory trials), with equal guess and lapse rates (at most 0.1) (Wichmann and Hill, 2001). The threshold was defined as the stimulus magnitude at the 50% point. Changes in threshold were expressed as $\log_2$(During/Pre).

In the Number Simon task (Figure 1E, F; PsychoPy v2024.2.4), based on Fischer et al. (2018) and adapted from Fischer et al. (2008), a digit between 1 and 9 (excluding 5) appeared to the left or right of fixation for 200 ms. Participants pressed a left key for digits below 5 and a right key for digits above 5 regardless of the digit's position. A trial was congruent (C) when the digit appeared on the same side as the required response key and incongruent (I) otherwise. The task comprised 192 trials in three blocks of 64, with each digit presented four times per congruency condition in each block in a randomized order. Responses were accepted for up to 1400 ms after digit offset, and reaction time (RT) was measured from digit offset; key presses made while the digit was shown were not registered. Errors and missing responses were followed by red feedback text, and the next trial began 3000 ms after digit offset. Only complete 192-trial runs were analyzed. In four runs (three participants, all in sham sessions), the task was restarted after an incomplete attempt, which was discarded, and only the subsequent complete run was analyzed. Correct trials were classified by the congruency of the previous trial (lowercase: c, i) and the current trial (uppercase: C, I), yielding four sequences (cC, cI, iC, iI), regardless of whether the previous trial was correct. The first trial of each block was excluded from sequence analyses. Trials without a response were counted as errors, and RT analyses included all correct trials without trimming. Five outcomes were prespecified: the Simon RT effect (mean RT on I minus C trials), the Simon error effect (error rate on I minus C trials), overall accuracy (proportion of correct responses across all trials), conflict adaptation (the congruency sequence effect in RT, CSE = (cI − cC) − (iI − iC); Gratton et al., 1992; Botvinick et al., 2001) and the incongruent-repetition benefit (mean RT on cI minus iI trials).

**2.10 Statistics**

For each outcome measure, the effect of interest was the difference-in-differences dNet = $(\text{taVNS}_t - \text{taVNS}_{\text{Pre}}) - (\text{Sham}_t - \text{Sham}_{\text{Pre}})$, where $t$ denotes Early, Mid or Post for the click measures and During for the behavioral measures. The statistical approach depended on the number of time points. All $p$ values were obtained from exact permutation tests. The significance level was set at $\alpha = .05$. Effects are reported as mean ± SEM, with Cohen's $d_z$ for dNet and partial eta squared ($\eta_p^2$) for ANOVA effects. Statistical analyses were performed in Python 3.14 with MNE-Python 1.13.2 (Gramfort et al., 2013), SciPy 1.18.1, statsmodels 0.15.0 and NumPy 2.5.3.

For the click-evoked potentials, changes from Pre at Early, Mid and Post were analyzed with a permutation-based 2 (condition) × 3 (time) repeated-measures ANOVA in the ten participants with complete EEG data. The condition main effect was tested using a sign-flip permutation test of the time-averaged dNet (Nichols and Holmes, 2002). The time main effect and Condition × Time interaction were tested by permuting the time labels within participants over all $6^9$ arrangements (the six orderings of the three time points in each participant, with the first participant held fixed), using the repeated-measures F statistic. When the Condition × Time interaction was significant, the changes from Pre at each time point were compared between conditions with sign-flip permutation tests of the dNet with Holm correction across the three time points (Holm, 1979).

For theta and behavioral measures, changes from Pre to Post (theta) or During (behavior) were compared between conditions using two-sided sign-flip permutation tests of dNet over all $2^N$ sign patterns (Nichols and Holmes, 2002), equivalent to testing the Condition × Time interaction in a 2 × 2 design. For the Simon task, the Benjamini–Hochberg procedure was applied within two families (Benjamini and Hochberg, 1995): the five Simon outcomes and the four congruency-sequence cells (cC, cI, iC, iI).

Time courses, time–frequency maps and scalp maps of the Post dNet were tested using sign-flip cluster-based permutation tests (Maris and Oostenveld, 2007), implemented in MNE-Python (permutation_cluster_1samp_test) (Gramfort et al., 2013). All $2^N$ sign patterns were evaluated (1024 for $N = 10$), with a two-sided cluster-forming threshold of $p < .05$ and cluster mass as the cluster statistic. The tests were applied to theta ITPC and ERSP time courses (0–500 ms), time–frequency maps (2–70 Hz, 0–500 ms) over the ROI, and 32-electrode maps of theta ITPC and ERSP (5–8 Hz, 0–300 ms), P0 (maximum at 5–15 ms), Pa–Nb (peak-to-trough amplitude) and P2 (mean at 125–175 ms). Electrode adjacency was defined by Delaunay triangulation of the standard 10–20 electrode positions (MNE-Python find_ch_adjacency). Significant clusters indicate effects at the cluster level without establishing their precise extent in time, frequency or space (Sassenhagen and Draschkow, 2019).

The exploratory correlation between theta ITPC dNet over the ROI at Post and conflict-adaptation dNet was assessed using Spearman's $\rho$, with an exact permutation test over all $N!$ pairings.

## 3 Results

### 3.1 Auditory-Evoked Potentials

During stimulation, a 25-Hz stimulation artifact was present in the raw EEG and was most prominent near the electrically stimulated left ear (Supplementary Figure S1A). At Cz, the spectral peak was more prominent under taVNS than under Sham at Mid (taVNS − Sham, +8.6 dB, $p = .034$; Early +4.0 dB, $p = .14$; Supplementary Figure S1B). Preprocessing reduced the peak to approximately the background level, with no significant difference between conditions at any time point (all $p \geq .06$; Supplementary Figure S1C). In ± averages designed to cancel click-locked responses, the residual 25-Hz amplitude was approximately 0.02 µV and did not differ significantly between conditions at any time point (all $p > .25$).

In both conditions, clicks evoked MLRs comprising P0 and the Pa–Nb complex at Cz and the left ROI (Figure 2A, B; Supplementary Figure S2). At Cz, the change in P0 from Pre showed a main effect of condition ($F(1, 9) = 4.74$, $p = .027$, $\eta_p^2 = .34$), but no main effect of time ($F(2, 18) = 0.44$, $p = .66$) or Condition × Time interaction ($F(2, 18) = 2.05$, $p = .16$; Figure 2C). Under Sham, P0 at Early, Mid, and Post was 0.03–0.09 µV below the Pre level, whereas under taVNS it remained close to the Pre level (0.00 to +0.10 µV). The scalp map of the Post P0 dNet showed a positive cluster comprising Fz, FC1, FC2, and Cz ($p = .018$; Figure 2D).

Over the left ROI, the change in Pa–Nb from Pre showed a Condition × Time interaction ($F(2, 18) = 5.45$, $p = .008$, $\eta_p^2 = .38$), but no main effect of condition ($F(1, 9) = 2.96$, $p = .12$) or time ($F(2, 18) = 0.04$, $p = .97$; Figure 2C). During Early, the change from Pre was +0.43 µV under taVNS and −0.10 µV under Sham, giving a dNet of +0.52 µV ($p = .041$). In the grand averages, this difference appeared as a larger Pa peak at about 30 ms and a deeper Nb trough at about 40 ms during Early than at Pre under taVNS (Figure 2B). At Mid and Post, dNet did not differ significantly from zero (Mid +0.40 µV, Post −0.20 µV; both $p = .24$). No significant cluster was detected in the scalp map of the Post Pa–Nb dNet (Figure 2D). As additional check for residual stimulation artifacts, Pa–Nb was also measured in the ± averages. The Early condition difference over the left ROI was +0.04 µV ($p = .65$), less than one-tenth of the Early dNet.

Clicks also evoked a slow-vertex P2 peaking at approximately 155 ms over the ROI (Figure 3A). The change in P2 from Pre showed a main effect of condition over the ROI ($F(1, 9) = 6.23$, $p = .029$, $\eta_p^2 = .41$) but no main effect of time or Condition × Time interaction (both $p \geq .25$; Figure 3B). From Pre to Post, P2 over the ROI increased by 0.22 µV under Sham and decreased by 0.13 µV under taVNS. The scalp map of the Post dNet of mean P2 amplitude showed a negative cluster comprising Fz, F3, FC1, FC2, and Cz ($p = .027$; Figure 3C).

**3.2 Theta Phase Locking and Power**

Theta measures were compared between Pre and Post ($N = 10$; Figure 4). The change in theta ITPC over the ROI from Pre to Post was +0.031 under Sham and −0.030 under taVNS, giving a dNet of −0.062 ± 0.019 ($p = .018$, $d_z = -1.02$; Figure 4C). The dNet time course showed a significant cluster spanning 0–420 ms ($p = .016$), and the time–frequency map showed a cluster spanning 4.4–13.1 Hz and 0–460 ms ($p = .016$; Figure 4A), with the strongest effect near 6 Hz. The scalp map showed a negative cluster comprising Fz, FC1, FC2, Cz, CP1, CP2, and Fp2 ($p = .025$; Figure 4E). The ITPC reduction was also detected at Cz using a 1-s discrete Fourier transform (5–8 Hz; −0.046 ± 0.013, $p = .012$, $d_z = -1.16$; Supplementary Figure S3).

The change in theta ERSP over the ROI from Pre to Post was −0.15 dB under Sham and −0.14 dB under taVNS, giving a dNet of +0.01 ± 0.14 dB that was not significant ($p = .94$, $d_z = +0.02$; Figure 4D). No significant clusters were detected in the ERSP dNet time course (no time point exceeded the cluster-forming threshold), time–frequency map ($p \geq .82$), or scalp map ($p$

≥ .23; Figure 4B, F). Scalp maps of theta ITPC and ERSP for each condition and time point are shown in Supplementary Figure S4.

**3.3 Behavioral Performance**

Changes in sensory discrimination thresholds did not differ significantly between conditions ($N = 11$; visual orientation: $\log_2$-ratio dNet +0.39 ± 0.35, $p = .28$, $d_z = 0.34$; auditory frequency: −0.21 ± 0.28, $p = .47$, $d_z = -0.22$; Supplementary Figure S5).

In the Simon task ($N = 9$), conflict adaptation changed from Pre to During by −26.4 ms under Sham and +31.9 ms under taVNS (Figure 5A), giving a dNet of +58.2 ± 16.7 ms ($p = .004$, $d_z = 1.16$). The dNet was positive in all nine participants and remained significant after Benjamini–Hochberg correction across the five prespecified Simon outcomes ($q = .020$). The incongruent-repetition benefit changed in the same direction (dNet +34.4 ± 13.4 ms, $p = .043$, $d_z = 0.86$), but did not survive the correction ($q = .11$; Figure 5B). No significant differences were found for the Simon RT effect (dNet −12.0 ± 7.4 ms, $p = .18$, $q = .30$; Figure 5C), Simon error effect ($p = .97$, $q = .97$), or overall accuracy ($p = .91$, $q = .97$; Figure 5D). The RT changes in the four congruency-sequence cells did not differ significantly between conditions after correction (all $q \geq .09$; Supplementary Figure S6). In an exploratory analysis of participants with both EEG and Simon task data ($N = 8$), theta ITPC dNet over the ROI at Post correlated negatively with conflict-adaptation dNet (Spearman's $\rho = -.76$, $p = .037$; Figure 5E). Greater decreases in theta phase locking were associated with greater increases in conflict adaptation.

## 4 Discussion

This study tested whether taVNS changes human auditory responses in the directions predicted by the FF/FB shift proposed on the basis of rat studies (Kumagai et al., 2025). Relative to Sham, the auditory measures changed as predicted. Among the predominantly afferent MLRs, P0 did not show the decline observed under Sham, and Pa–Nb increased during the first ten minutes of stimulation. Among the later responses, which may have greater FB contributions, P2 did not show the increase observed under Sham, and click-locked theta phase locking was lower after stimulation, without a significant between-condition difference in the change in theta power. Behaviorally, conflict adaptation increased, whereas sensory thresholds and the overall Simon effect did not differ significantly between conditions.

**4.1 Early and Late Auditory Responses**

P0 and Pa–Nb, which predominantly reflect bottom-up transmission (Garrido et al., 2007), changed as predicted for FF-dominated responses. This FF interpretation is supported by rat studies linking early evoked responses to basic acoustic feature encoding and the organization of ascending auditory pathways (Takahashi et al., 2004, 2005a, 2005b; Shiramatsu et al., 2016). Under Sham, P0 declined from its baseline over the session, possibly because the repeated clicks became predictable and their afferent signal was down-weighted (Feldman and Friston, 2010). Under taVNS, P0 remained close to baseline, suggesting that taVNS preserved the gain of the afferent input rather than raising it above its initial level. Pa–Nb increased above baseline early during stimulation, consistent with the rapid enhancement of somatosensory thalamic transmission by VNS in the rat (Rodenkirch and Wang, 2020). The significant Condition × Time interaction indicates that this effect varied over the session, possibly reflecting cortical adaptation to the increased input during prolonged stimulation. These changes parallel the enhancement of superficial-layer responses and gamma activity by VNS in the rat auditory cortex (Takahashi et al., 2020; Kumagai et al., 2023).

P2, a later component likely to have a greater FB contribution than the MLRs (Garrido et al., 2007), increased across the Sham session but not under taVNS. Click-locked theta phase locking was also reduced after stimulation. In rats, VNS reduces auditory-evoked theta power through noradrenergic mechanisms (Kumagai et al., 2023), suggesting that noradrenergic modulation may contribute to the theta response changes observed here. These responses to task-irrelevant clicks should be distinguished from frontal midline theta measured during cognitive-control tasks (Cavanagh and Frank, 2014).

The reduction in theta ITPC was not accompanied by a significant between-condition difference in the change in theta ERSP. ITPC measures phase consistency across trials, whereas ERSP measures changes in spectral power relative to baseline. These aspects of auditory responses can exhibit different dynamics: in the rat auditory cortex, phase-locking and amplitude adaptation differ in their frequency dependence and time course (Noda et al., 2014). Phase locking also reflects regularities across tone sequences, with temporal and spatial characteristics distinct from those of evoked-response amplitudes (Noda et al., 2017). The present ITPC reduction therefore indicates altered temporal coordination of responses to repetitive clicks, without a detectable corresponding change in theta power.

### 4.2 Interpretation in an FF/FB Framework

A key feature of the present study is the combined assessment of MLRs and later auditory responses. MLRs provide a complementary measure of predominantly ascending FF processing (Takahashi et al., 2004, 2005a, 2005b, 2007; Shiramatsu et al., 2016), allowing changes in later responses to be evaluated in relation to early sensory processing. Immediately after stimulation, changes from baseline were more positive for P0 and more negative for P2 and theta phase locking under taVNS than under Sham, with significant clusters over overlapping fronto-central sensors. The opposite directions of these changes ordered across processing stages are difficult to explain by a simple uniform change in response gain and may reflect a change in the relative weighting of FF and FB transmission.

Layer-resolved recordings in rats provide complementary evidence that VNS alters the cortical representation of auditory information. Sound frequency can be decoded from spatial patterns of band-specific power and inter-site phase locking during sustained auditory activity, with more accurate high-gamma representations in layers 4 and 5/6 than in layers 2/3 (Shiramatsu et al., 2019). VNS modifies these representations in a layer- and frequency-dependent manner: high-gamma decoding accuracy decreases in layers 5/6, with more limited improvements in layers 2/3 and 4, whereas theta-power-based decoding improves in layers 5/6 (Shiramatsu et al., 2025). These findings suggest that VNS reorganizes sensory representations across cortical layers and frequency bands. Such reorganization provides a circuit-level context for the opposing changes in early and late auditory responses observed here and is compatible with the proposed reweighting of FF and FB processing. Taken together, the present results extend the pattern predicted by the FF-over-FB hypothesis from invasive VNS studies in rats to taVNS in humans.

### 4.3 Conflict Adaptation and the Precision of Prediction Errors

Under taVNS, conflict adaptation increased without a significant change in sensory thresholds, the Simon effect or overall accuracy. Within a predictive coding framework, a mismatch between predicted and actual conflict can be interpreted as a prediction error at the level of cognitive control (Gratton et al., 1992; Botvinick et al., 2001; Jiang et al., 2014), so taVNS may have strengthened the control adjustments driven by this error. This is consistent with the prediction that VNS benefits tasks requiring updates to internal models (Kumagai et al., 2025) and with the enhancement of conflict adaptation reported for taVNS (Fischer et al., 2018). In an exploratory analysis, greater decreases in theta phase locking were associated with greater increases in conflict adaptation across participants (Spearman's $\rho = -.76$, $N = 8$). Although this association does not establish a shared mechanism, we propose a common neuromodulatory explanation for these changes: the increased precision weighting of prediction errors relative to predictions, potentially mediated by acetylcholine and noradrenaline (Yu and Dayan, 2005; Feldman and Friston, 2010).

### 4.4 Limitations

Several limitations qualify these interpretations. The study was not preregistered, and the small sample sizes ($N$ = 8–11) limit the precision of the estimates, particularly for the exploratory correlation. The single-blind design does not exclude experimenter-related bias, and differences in perceived stimulation between conditions could have compromised participant blinding. Residual stimulator and muscle artifacts cannot be excluded completely, and theta effects were assessed only as Pre-to-Post changes, leaving their time course during stimulation unknown. The present EEG analysis did not directly assess the direction of FF/FB signaling, so the proposed shift in FF/FB processing remains an interpretation rather than a demonstrated mechanism.

## 5 Conclusion

Relative to Sham, transcutaneous auricular VNS (taVNS) preserved or enhanced early MLRs, attenuated later auditory responses, and increased conflict adaptation. These findings are consistent with, but do not demonstrate, a shift toward FF relative to FB processing. Direct measures of directional FF/FB signaling and computational tests of prediction-error processing are needed to further evaluate this interpretation.

## Author contributions

Issey Kamatani: Methodology, Investigation, Data curation, Formal analysis, Writing – original draft, and Visualization. Yasuo Kabe: Methodology, Investigation, Validation, and Writing – review & editing. Shinichi Kumagai: Methodology, Formal analysis, Writing – original draft, Writing – review & editing and Visualization. Yoshiyuki Onuki: Methodology and Writing – review & editing. Hexin Xu: Investigation and Writing – review & editing. Tomoyo Isoguchi Shiramatsu: Methodology, Funding acquisition and Writing – review & editing. Naoto Kunii: Supervision, Funding acquisition and Writing – review & editing. Kensuke Kawai: Supervision, Funding acquisition and Writing – review & editing. Hirokazu Takahashi: Conceptualization, Supervision, Project administration, Funding acquisition, and Writing – review & editing.

## Funding

This work was supported by JSPS KAKENHI (23H03465, 23H04336, 23H03023, 23KJ1866, 24H01544, 25H02600, 25K22825, 26H02517), AMED (24wm0625401h0001), JST (JPMJPR22S8), NEDO (JPNP14004), the Asahi Glass Foundation and the SECOM Science and Technology Foundation.

**Conflict of Interest**

The authors declare that there are no conflicts of interest associated with this manuscript.

**Declaration of generative AI and AI-assisted technologies in the manuscript preparation process**

During the preparation of this work, the authors used Claude and ChatGPT to improve language, readability, and manuscript organization. After using these tools, the authors reviewed and edited the content as needed and take full responsibility for the content of the published article.

**Data statement**

The data supporting the findings of this study will be made publicly available in the Zenodo repository upon publication of this article.

## Figures

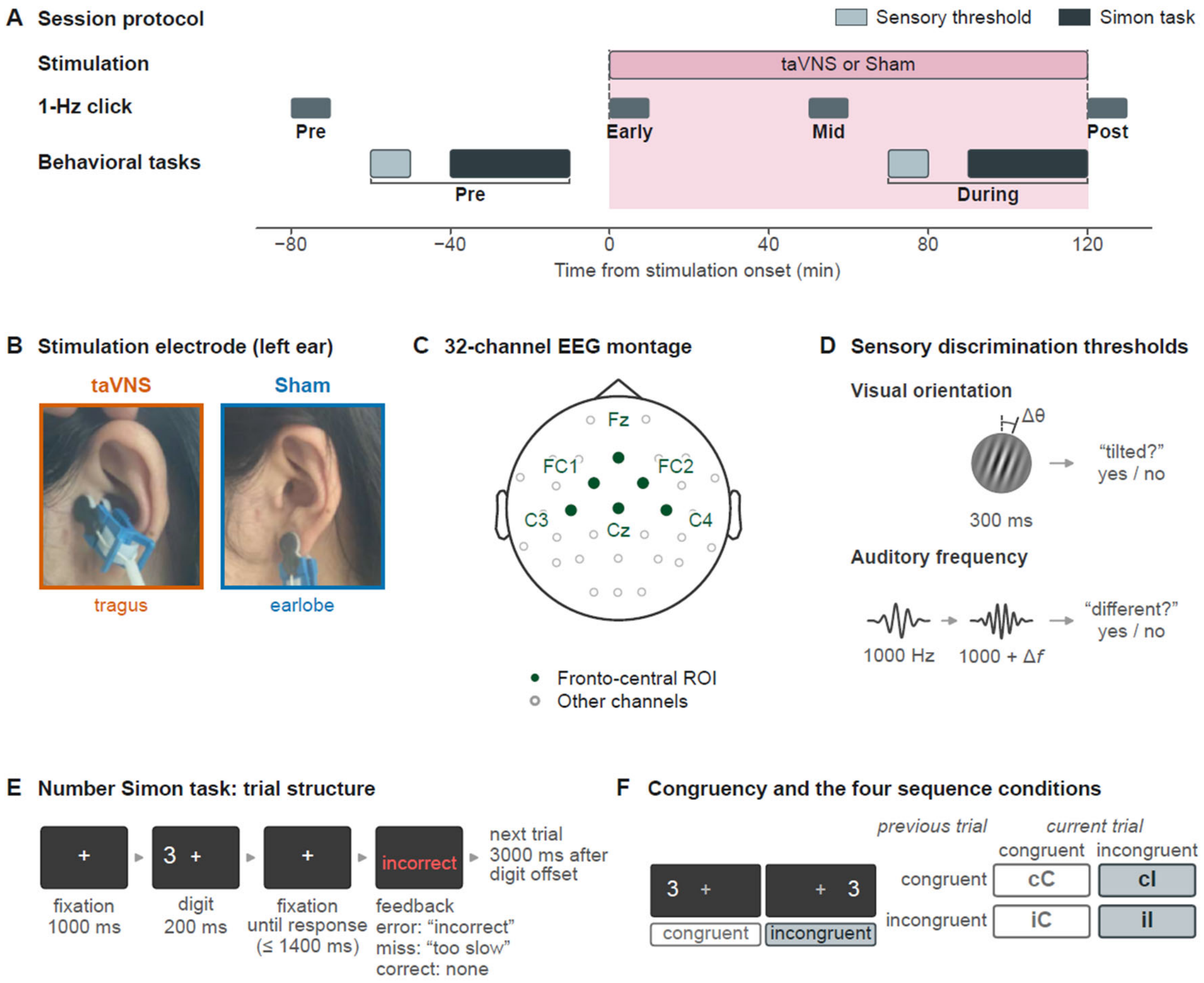

**FIGURE 1** Study design. (A) Session protocol, identical in the taVNS and Sham sessions. Four 1-Hz click blocks (Pre, Early, Mid, Post) were recorded, and the behavioral tasks (sensory thresholds and the Number Simon task) were administered before (Pre) and during (During) stimulation. The pink band marks the stimulation period. (B) Stimulation sites (photographs): left tragus (taVNS) and left earlobe (Sham). (C) EEG montage (32 channels) showing the fronto-central region of interest (ROI: Fz, FC1, FC2, Cz, C3 and C4; filled). (D) Visual orientation and auditory frequency discrimination tasks. (E) Number Simon trial structure. The digit was shown for 200 ms, responses were accepted for up to 1400 ms after its offset, and errors and missing responses were followed by red feedback text. Participants pressed a left key for digits below 5 and a right key for digits above 5, regardless of the digit's position. (F) Congruency and the four sequence conditions. A trial was congruent when the digit appeared on the same side as the required response key (e.g., 3 on the left) and incongruent otherwise. Lowercase and uppercase letters denote the previous and current trials, respectively. The measures were calculated from the mean reaction time (RT) in each condition: Simon RT effect = I − C; congruency sequence effect (CSE, an index of conflict adaptation) = (cI − cC) − (iI − iC);

incongruent-repetition benefit = cI − iI.

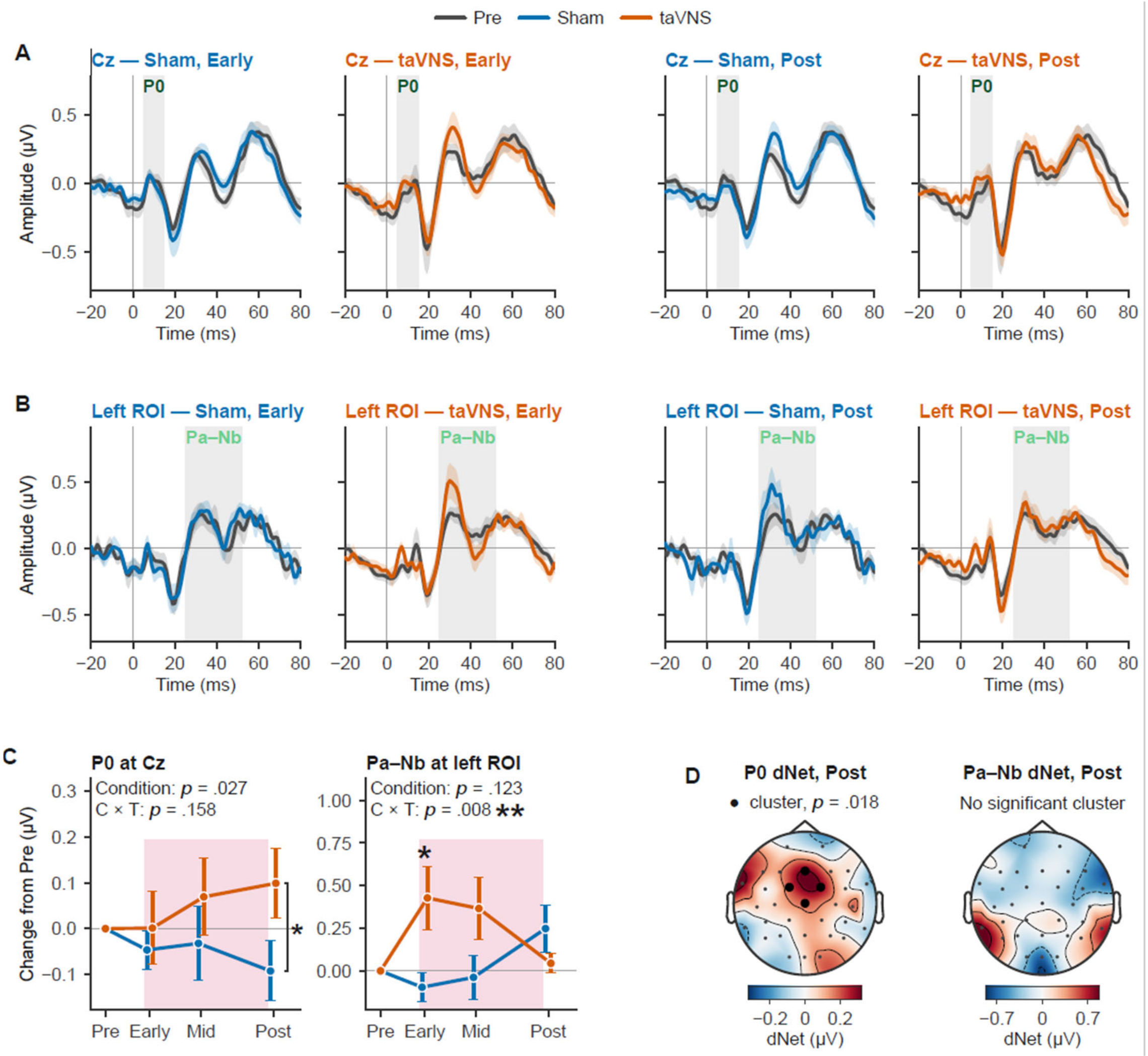

**FIGURE 2** MLRs across the stimulation session: P0 at Cz and Pa–Nb over the left ROI. (A) Grand-average MLRs (N = 10) at Cz at Pre (gray) and at either Early (left pair) or Post (right pair); Sham (blue) and taVNS (orange) are shown in separate panels using the same y-axis scale. Shading indicates ± SEM. The gray band marks the P0 analysis window (5–15 ms). (B) As in (A), over the left ROI (FC1, C3); the gray band marks the Pa–Nb analysis window (25–52 ms). (C) Changes from Pre in P0 at Cz and Pa–Nb over the left ROI, mean ± SEM. The pink band marks the stimulation period. Early and Mid were recorded during stimulation and Post immediately after stimulation ended (Figure 1A). In (C), $p$ values are from the permutation-based 2 (condition) × 3 (time) repeated-measures ANOVA of changes from Pre ($N = 10$). The bracket to the right of the P0 plot indicates a main effect of condition, C × T indicates the Condition × Time interaction, and the asterisk above Early indicates a significant dNet after Holm correction (*$p < .05$, **$p < .01$). (D) Scalp maps of the Post dNet for P0 and Pa–Nb, calculated as ($taVNS_{Post} - taVNS_{Pre}$) − ($Sham_{Post} - Sham_{Pre}$) at each electrode ($N = 10$). Black dots

indicate electrodes belonging to a significant cluster (sign-flip cluster-based permutation test, cluster-level $p < .05$).

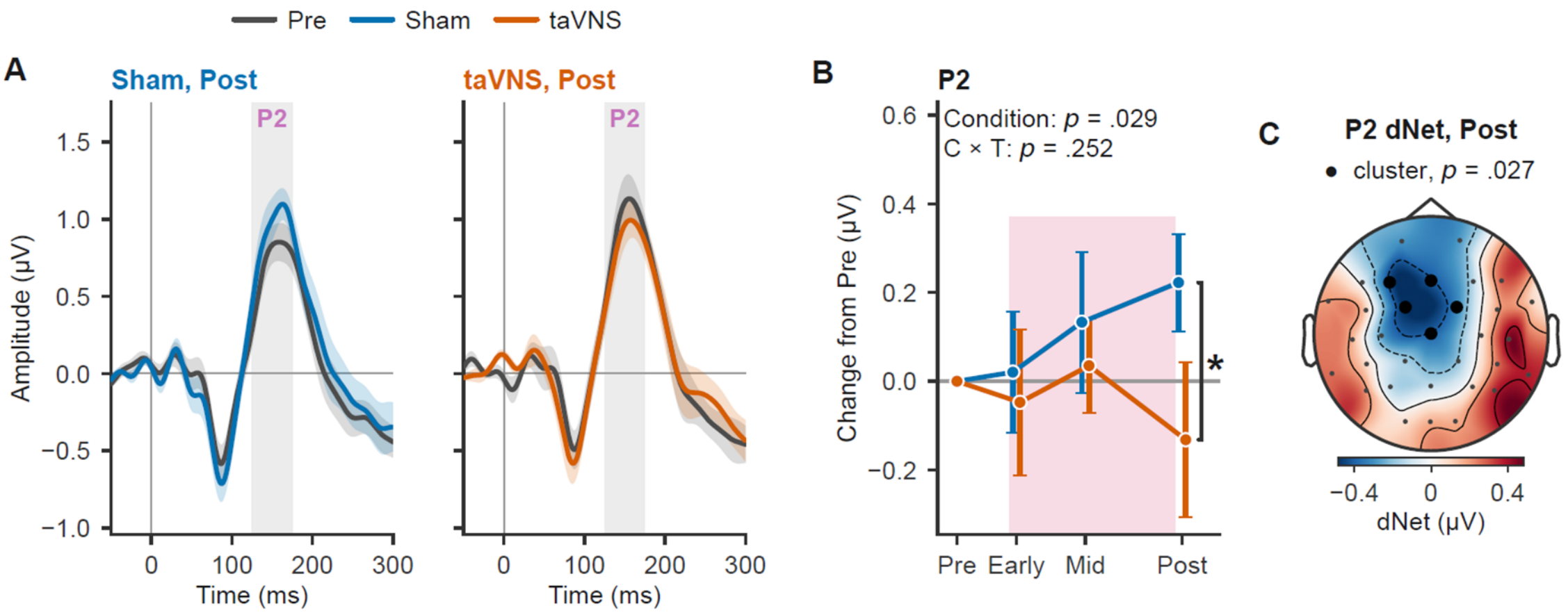

**FIGURE 3** Slow-vertex P2 over the fronto-central ROI across the stimulation session. (A) Grand-average slow-vertex responses ($N = 10$) over the ROI at Pre (gray) and Post (color), with Sham and taVNS shown in separate panels using the same y-axis scale. Shading indicates ± SEM. The gray band marks the P2 analysis window (125–175 ms). (B) Change from Pre in P2 over the ROI, as in Figure 2C; the bracket indicates the main effect of condition ($p < .05$). (C) Scalp map of the Post dNet for the mean P2 amplitude (125–175 ms; $N = 10$). Black dots indicate electrodes belonging to a significant cluster (sign-flip cluster-based permutation test, cluster-level $p < .05$).

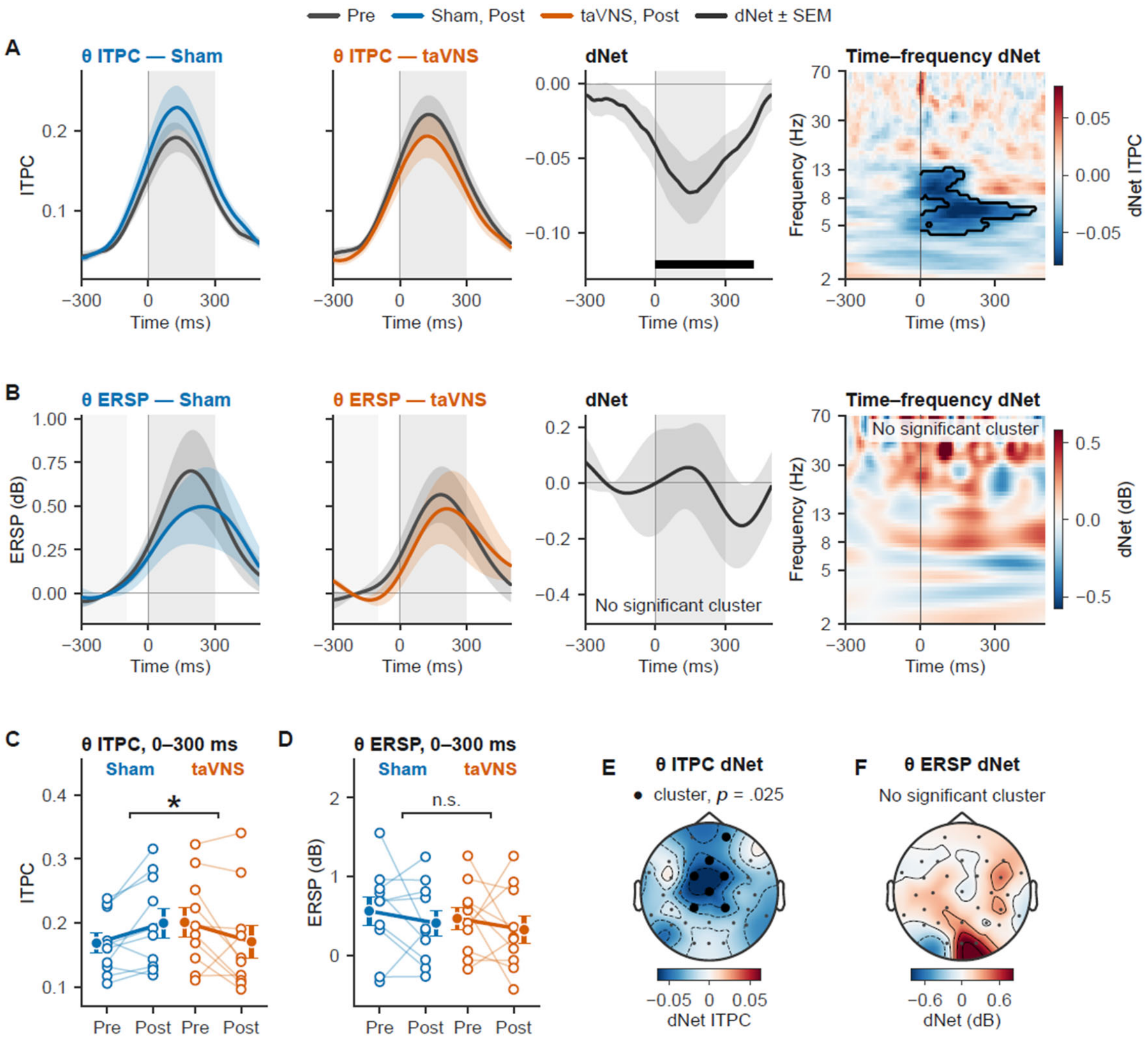

**FIGURE 4** Theta phase locking and power after stimulation. All panels show Pre-to-Post comparison or the corresponding dNet ($N = 10$); dNet = ($taVNS_{Post} - taVNS_{Pre}$) − ($Sham_{Post} - Sham_{Pre}$). Panels A–D show measures calculated from signals averaged across the fronto-central ROI electrodes. (A) Theta (5–8 Hz) inter-trial phase coherence (ITPC): time courses for Sham and taVNS (Pre, gray; Post, color; shading, ± SEM), the dNet time course (mean ± SEM), and the dNet time–frequency map. The black bar and the black outline indicate significant clusters (cluster-level $p < .05$). The gray band (0–300 ms) marks the analysis window used in (C–F). (B) As in (A), for theta event-related spectral perturbation (ERSP), expressed in dB relative to the baseline from −300 to −100 ms (light gray band). (C, D) Theta ITPC and ERSP (0–300 ms) at Pre and Post: individual participants (open circles) and mean ± SEM (filled circles and error bars). Brackets indicate two-sided sign-flip permutation tests of dNet (*$p < .05$; n.s., not significant). (E, F) Scalp maps of theta ITPC and ERSP dNet (0–300 ms). Black dots indicate electrodes belonging to a significant cluster (cluster-level $p < .05$).

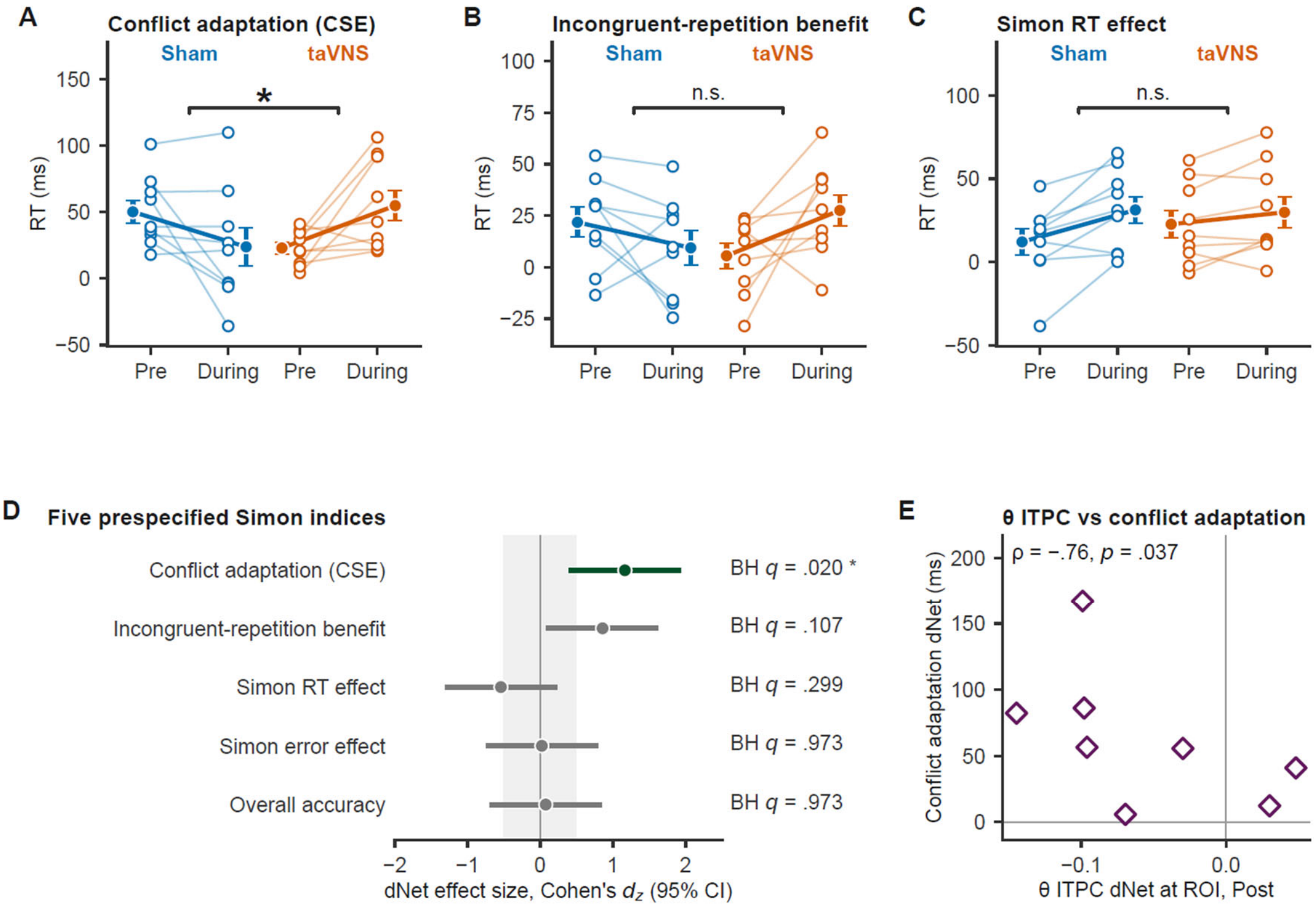

**FIGURE 5** Simon task performance and its relation to theta phase locking. Pre, before stimulation; During, during stimulation; dNet, ($taVNS_{During}$ − $taVNS_{Pre}$) − ($Sham_{During}$ − $Sham_{Pre}$) for behavioral measures. (A–C) Conflict adaptation (congruency sequence effect, CSE), incongruent-repetition benefit, and Simon RT effect at Pre and During under Sham and taVNS. Open circles represent individual participants ($N = 9$); filled circles and error bars indicate the mean ± SEM. Brackets indicate two-sided sign-flip permutation tests of dNet, with Benjamini–Hochberg correction across the five outcomes in (D): *$q$ < .05; n.s., not significant. (D) Effect sizes for dNet (Cohen's $d_z$ with 95% confidence intervals) for the five Simon outcomes. Green indicates $q < .05$, and the gray band indicates $|d_z| < 0.5$. (E) Exploratory association between theta ITPC dNet over the ROI at Post and conflict-adaptation dNet (Spearman's $\rho$, exact permutation $p$; $N = 8$).

**Supplementary Materials for**

# Transcutaneous auricular vagus nerve stimulation modulates auditory-evoked responses and conflict adaptation in humans

Issey Kamatani[1], Yasuo Kabe[1], Shinichi Kumagai[1,2], Yoshiyuki Onuki[2], Hexin Xu[1], Tomoyo Isoguchi Shiramatsu[1], Naoto Kunii[2], Kensuke Kawai[2] and Hirokazu Takahashi[1*]

[1]Graduate School of Information Science and Technology, The University of Tokyo, Tokyo, Japan; [2]Department of Neurosurgery, Jichi Medical University, Shimotsuke, Japan

Correspondence to: Hirokazu Takahashi (takahashi@i.u-tokyo.ac.jp)

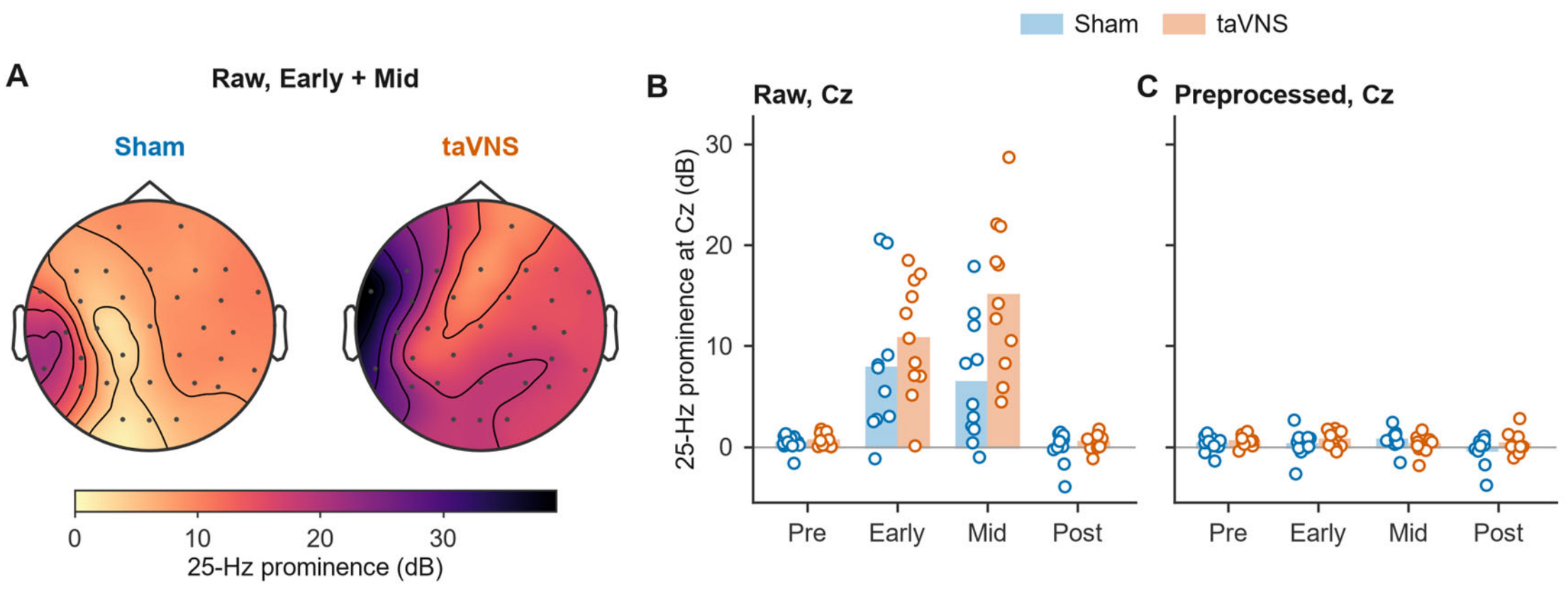

**Supplementary Figure S1.** 25-Hz stimulation artifact before and after preprocessing. Prominence (dB) of the 25-Hz stimulator line relative to the local spectral background in the four 1-Hz click recordings: Pre (before stimulation), Early and Mid (during stimulation) and Post (immediately after stimulation ended), as shown in Figure 1A. The central 180 s of each recording was analyzed using Welch's method (4-s Hann windows, 50% overlap, median averaging). Peak prominence was calculated as $10 \cdot \log_{10}$(peak/background), where peak was the maximum power within ±0.3 Hz of 25 Hz and background was the median power 1.5–6 Hz away from the peak on either side. (A) Scalp maps of the prominence in the raw recordings during stimulation (Early and Mid averaged within each participant) for Sham and taVNS, shown on a common color scale. Prominence was highest over the left temporal electrodes, near the stimulated left ear. (B) Raw recordings at Cz (common average reference). During stimulation, the 25-Hz peak was present in both conditions. The taVNS − Sham difference was +4.0 dB at Early ($p$ = .14) and +8.6 dB at Mid ($p$ = .034). At Pre and Post, mean prominence was near background level in both conditions (between-condition comparisons, $p \geq .18$). (C) Preprocessed recordings at Cz, plotted on the same scale as (B). Mean prominence was below 1 dB at every time point in both conditions, with no significant taVNS − Sham difference (all $p \geq .06$). In (B) and (C), bars show means, and open circles represent individual participants ($N$ = 10–11). Comparisons used two-sided sign-flip permutation tests without correction for multiple comparisons.

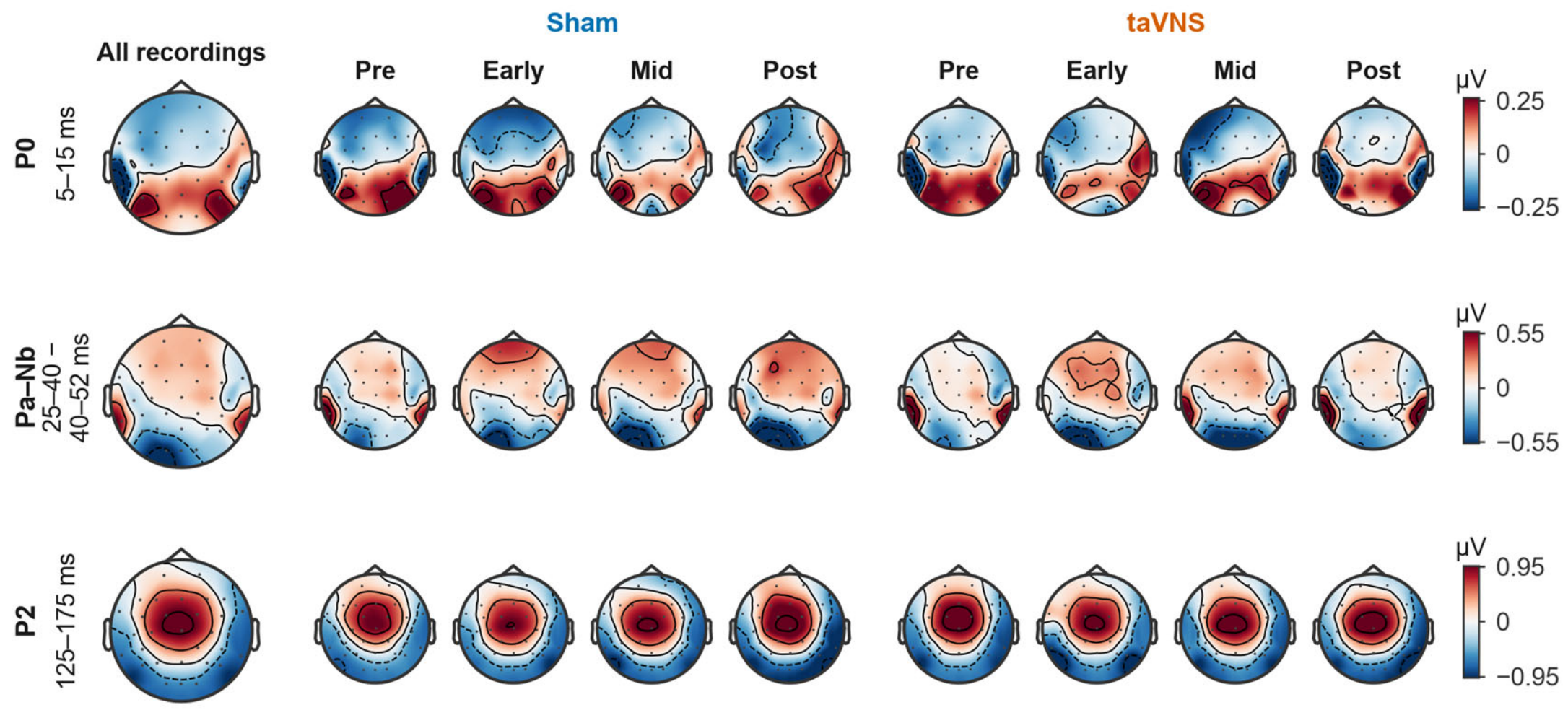

**Supplementary Figure S2.** Scalp distributions of P0, Pa–Nb, and P2. Maps show grand-average amplitudes at each electrode, without subtraction between conditions or time points. P0 (5–15 ms) and Pa–Nb (mean amplitude within 25–40 ms minus that within 40–52 ms) were obtained from the middle-latency responses (10–250 Hz), and P2 (125–175 ms) from the slow-vertex responses (1–30 Hz). All maps use the common-average reference. Left: averages across all recordings, computed first within each participant and then across participants ($N = 11$). Right: each condition and time point ($N = 10$–$11$). Each row uses a common, symmetric color scale with limits set at the 95th percentile of absolute amplitudes across its maps. No statistical tests were applied to these maps.

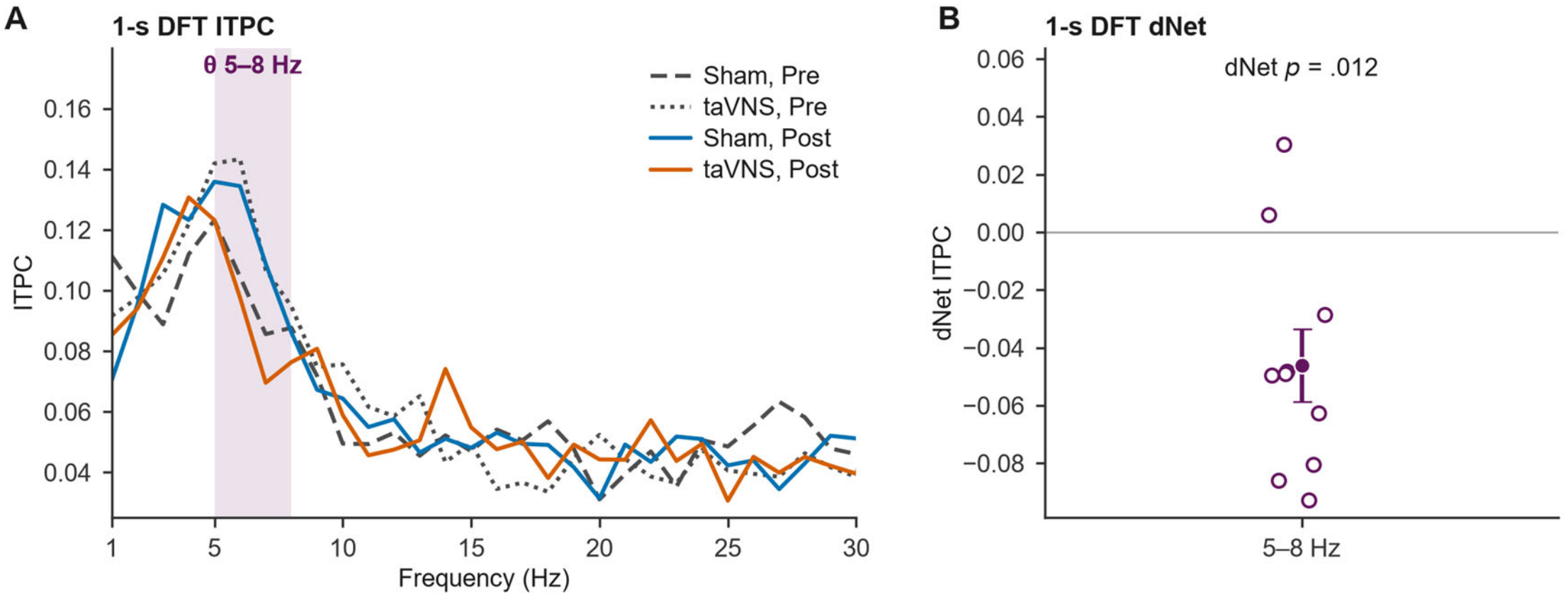

**Supplementary Figure S3.** Theta phase coherence estimated with a 1-s discrete Fourier transform. (A) Inter-trial phase coherence (ITPC) at Cz from 1 to 30 Hz, computed using a 1-s discrete Fourier transform of each inter-click interval, for Sham and taVNS at Pre (gray dashed line for Sham and gray dotted line for taVNS) and Post (colored lines) ($N = 10$). The shaded band marks the theta range (5–8 Hz). (B) Net change (dNet) in ITPC averaged over 5–8 Hz, calculated as $(\text{taVNS}_{\text{Post}} - \text{taVNS}_{\text{Pre}}) - (\text{Sham}_{\text{Post}} - \text{Sham}_{\text{Pre}})$. Open circles represent individual participants, and the filled circle and error bar indicate the mean ± SEM. The $p$ value is from a two-sided sign-flip permutation test.

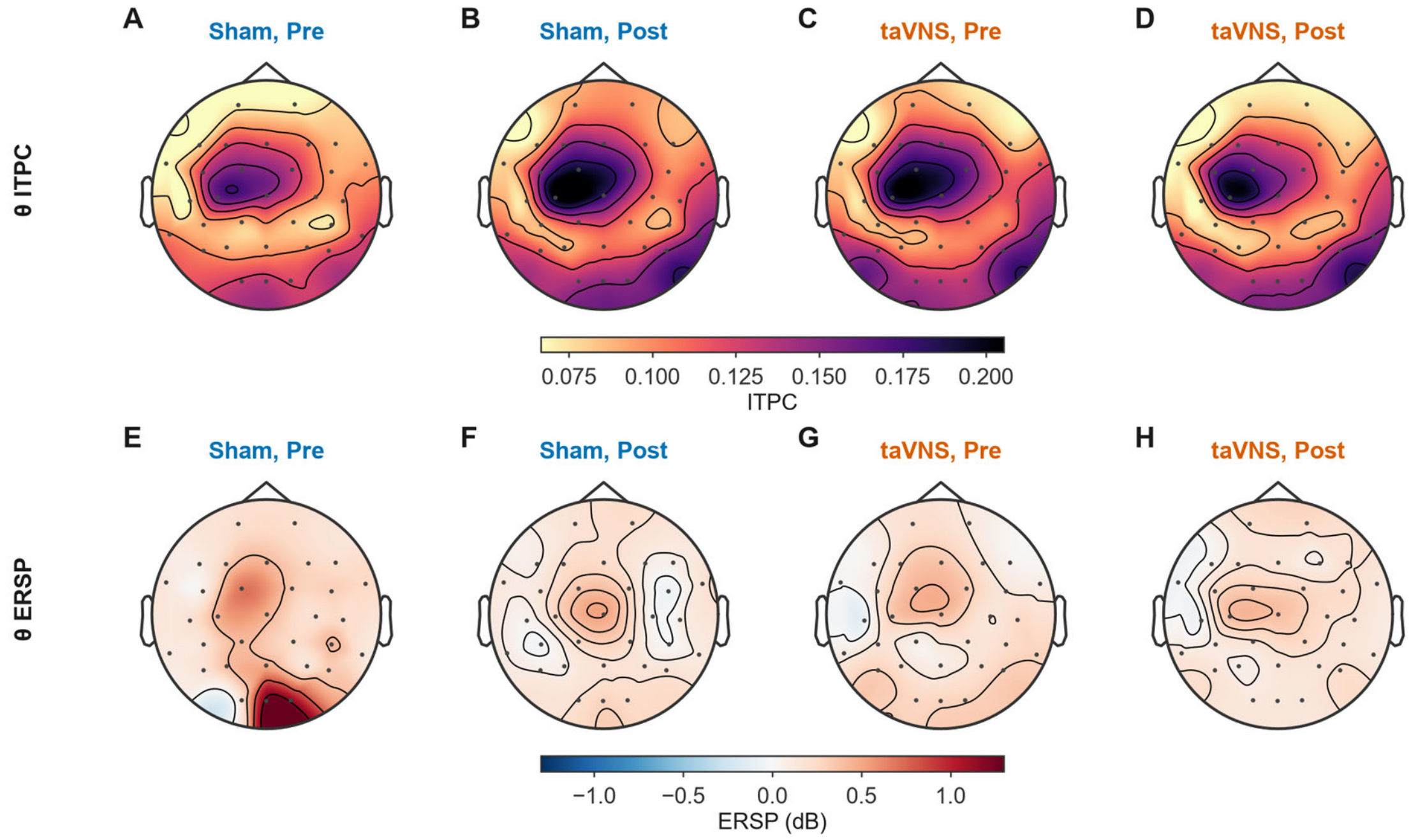

**Supplementary Figure S4.** Scalp maps of theta ITPC and ERSP. Theta (5–8 Hz) ITPC (A–D) and event-related spectral perturbation (ERSP) (E–H), averaged over 0–300 ms at Pre and Post for Sham and taVNS ($N = 10$). ERSP is expressed in dB relative to the baseline from −300 to −100 ms. Color scales are shared within each row. No statistical tests were applied to these maps. The corresponding dNet maps and cluster-based permutation tests are shown in Figure 4E, F.

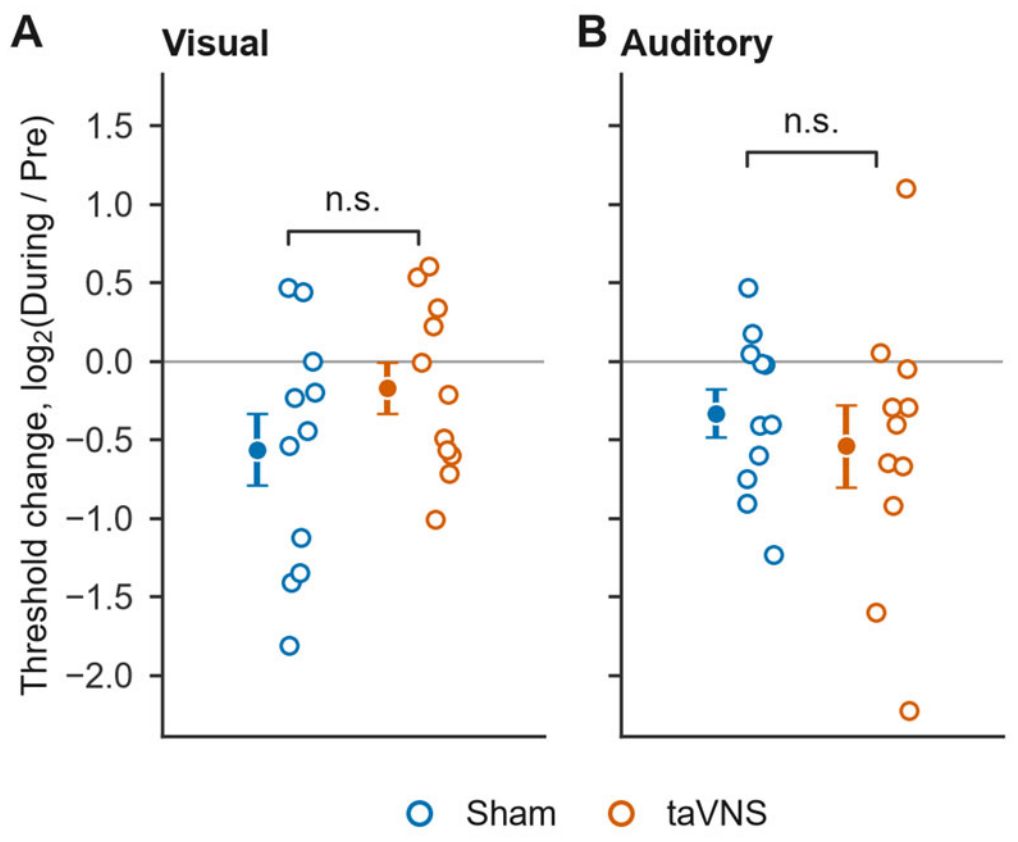

**Supplementary Figure S5.** Visual and auditory discrimination thresholds. Changes in the visual orientation (A) and auditory frequency (B) discrimination thresholds from Pre to During, expressed as $\log_2$(During/Pre). Zero denotes no change, and negative values indicate lower thresholds during stimulation than before. Open circles represent individual participants ($N = 11$); filled circles and error bars indicate the mean ± SEM. Brackets indicate two-sided sign-flip permutation tests of dNet, defined as the taVNS − Sham difference in the $\log_2$ ratios; n.s., not significant.

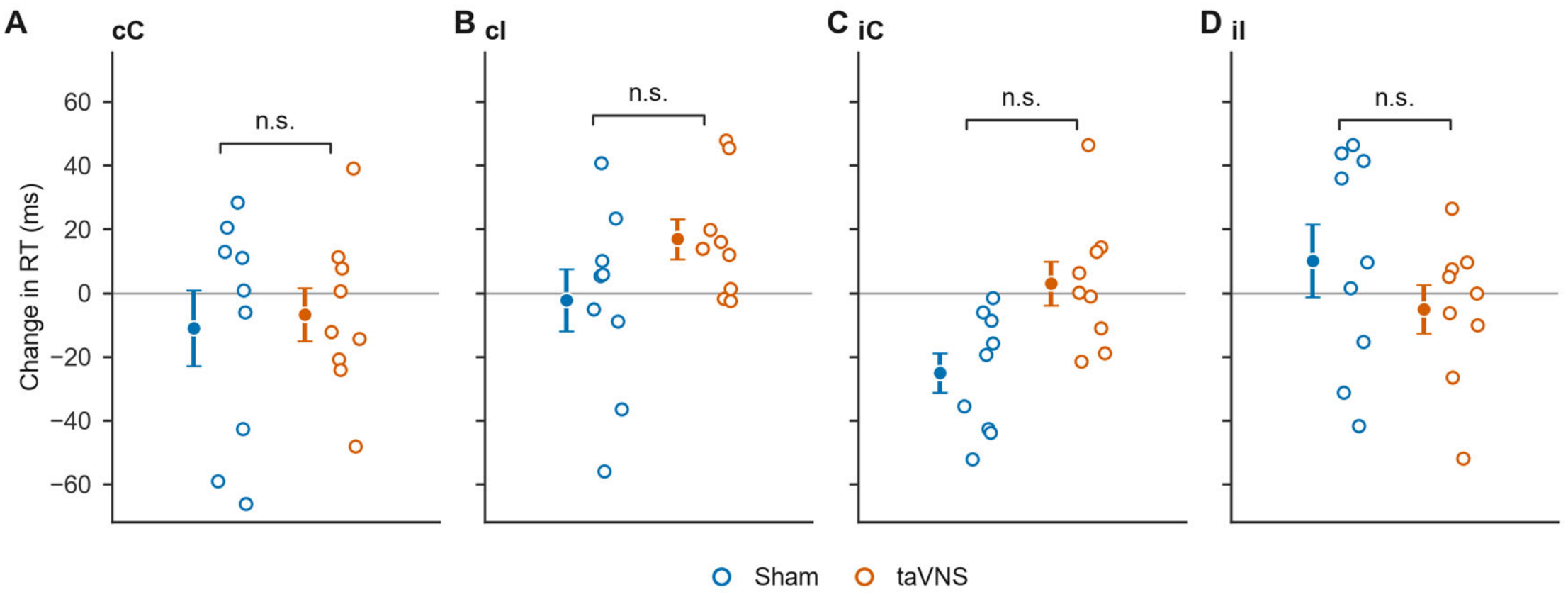

**Supplementary Figure S6.** Reaction-time changes in the four congruency sequences. Changes in reaction time (RT) from Pre to During in the congruent–congruent (cC; A), congruent–incongruent (cI; B), incongruent–congruent (iC; C) and incongruent–incongruent (iI; D) sequences under Sham and taVNS. Open circles represent individual participant ($N = 9$); filled circles and error bars indicate the mean ± SEM. Lowercase and uppercase letters denote the previous and current trials, respectively. Brackets indicate two-sided sign-flip permutation tests of $dNet = (taVNS_{During} - taVNS_{Pre}) - (Sham_{During} - Sham_{Pre})$, with Benjamini–Hochberg correction across the four cells; n.s., not significant.